\documentclass[final,5p,times,twocolumn]{elsarticle}
\usepackage{amssymb}

\journal{eTransportation}

\usepackage[cmex10]{amsmath}
\usepackage[utf8]{inputenc}
\usepackage{siunitx}
\usepackage{graphicx}
\usepackage{algpseudocode}
\usepackage[ruled]{algorithm2e}
\usepackage{dblfloatfix} 
\usepackage{float}

\SetKwInput{KwInput}{Input} 
\SetKwInput{KwOutput}{Output} 

\usepackage{tikz}
\usetikzlibrary{backgrounds}

\usepackage{xcolor} 

\usepackage{placeins}

\begin{document}

\begin{frontmatter}

\title{Energy Management of a Multi-Battery System for Renewable-Based\\High Power EV Charging}

\author{Jan Engelhardt}
\author{Jan Martin Zepter}
\author{Tatiana Gabderakhmanova}
\author{Mattia Marinelli}

\address{Department of Wind and Energy Systems\\Technical University of Denmark (DTU)\\Frederiksborgvej 399, 4000 Roskilde}

\begin{abstract}
Hybrid fast-charging stations with battery storage and local renewable generation can facilitate low-carbon electric vehicle (EV) charging, while reducing the stress on the distribution grid. This paper proposes energy management strategies for a novel multi-battery design that directly connects its strings to other DC components through a busbar matrix without the need for interfacing power converters. Hence, the energy management system has two degrees of control: (i) allocating strings to other DC microgrid components, in this case a photovoltaic system, two EV fast chargers, and a grid-tie inverter, and (ii) managing the energy exchange with the local distribution grid. For the grid exchange, a basic droop control is compared to an enhanced control including forecasts in the decision making. To this end, this paper evaluates results from multiple Monte Carlo simulations capturing the uncertainty of EV charging. For a realistic charging behaviour in each simulation run, random fast-charging profiles were created based on probability distributions of actual fast-charging data for arrival time, charging duration, and requested energy. The impact of different utilisation levels of the chargers was assessed by varying the average charging instances from 1 to 30 EVs per day. Using actual photovoltaic measurements from different months, the numerical analyses show that the enhanced control increases self-sufficiency by reducing grid exchange, and decreases the number of battery cycles. However, the enhanced control operates the battery closer to its charge limits, which may accelerate calendar ageing. 
\end{abstract}


\end{frontmatter}

\section{Introduction}

The deployment of fast-charging stations facilitate the wide-spread adoption of electric vehicles (EVs) as they address typical concerns such as range anxiety and long charging times \cite{Brand.2017, Zhang.2018b}. At the same time, the electrification of the transport sector challenges the operation of electric grids as EVs increase both the overall energy consumption and the peak load demand \cite{Yilmaz.2013,Calearo.2019}. In particular, fast charging stations with more than 50\,kW can stress distribution grids to their limits and are only feasible where these are sufficiently robust \cite{Mauri.2012,Dharmakeerthi.2014}. 

In the past years, there has been an increasing interest in equipping fast chargers with stationary battery systems that serve as a buffer during high power charging \cite{Mahfouz.2020}. The combination of EV chargers, batteries, and renewable energy sources (RES) in a hybrid system further allows to facilitate the local usage of renewable energy and make EV chargers to a certain degree self-sufficient \cite{Simpson.2012, Liu.2020}. Conventionally, these units are connected to a common bus via their respective power converters which are necessary to convert the voltage level between unit and bus, and to control the power flows in the hybrid system \cite{Khalid.2019,Sbordone.2015}. Several configurations of this design have been presented in literature. Vermaak et al. \cite{Vermaak.2014} propose a hybrid system comprising photovoltaic (PV), wind, and a stationary battery energy storage system (BESS) for charging electric vehicles in areas without grid connection. Cunha et al. \cite{Cunha.2016} focus their investigation on vanadium redox flow batteries to provide peak shaving capability in fast-charging stations. The value of energy storage in an electric bus fast charging station was highlighted by Ding et al \cite{Ding.2015}, showing that energy storage contributed to a total cost reduction by 22.85\,\%. Hafez et al. \cite{Hafez.2017} dimension a microgrid with integrated chargers, renewable energy sources, and grid connection to reduce life cycle costs for an assumed daily number of 20 EVs. The charging demand profile was regarded as fixed and based on logged drive cycles of a 2013 Chevrolet Volt with 16\,kWh battery capacity. However, for studying the impact of EV charging on the power system, the uncertainty and diversity of EV charging profiles should be considered \cite{Calearo.2021}. Domínguez-Navarro et al. \cite{DominguezNavarro.2019} address this point by including the uncertainty of EV demand. They applied a genetic algorithm to optimise the installation and operation of the EV fast-charging station.
Generally, the performance of hybrid systems depends on the energy management system (EMS) which is responsible for planning, monitoring, and controlling the power flow between the different units, as well as the energy level of present storage systems. Energy management strategies can consist of basic algorithms where the battery is recharged whenever excess power is available \cite{Vermaak.2014} or target specific objectives while meeting the EV demand, such as minimizing energy costs \cite{Le.2015,Tushar.2014,Chaudhari.2016}, reduce the grid impact \cite{GarciaTrivino.2016,Deng.2016,Mumtaz.2017}, as well as decreasing losses on system level \cite{Gamboa.2010} or in the grid \cite{Acha.2010}. 

The H2020 InsulaE project explores the benefits of a novel fast-charging station design for decarbonising the transportation sector. The core of this prototype is formed by an innovative battery system with reconfigurable topology providing the intrinsic capability to adapt its voltage level during operation \cite{Engelhardt.2020,Engelhardt.2021}. The battery consists of subsystems (strings) that can be directly coupled with other DC components through a busbar matrix, without the need of any power converters or a common DC bus. This makes the BESS the central entity of the hybrid system, responsible for controlling the power flow between its strings and other components: a PV system, two EV fast chargers, and a grid-tie inverter \cite{Gabderakhmanova.2020}. To fulfill EV charging with a high share of renewable energy, an advanced energy management system has to be designed that incorporates the (partly contradicting) requirements of all system components. While energy management systems for the conventional setup with a single bus have been previously presented in above-mentioned literature, these cannot be applied to the new system due to its fundamentally different layout. Since the components are not connected to the same bus but can be individually coupled through a busbar matrix, the energy management has the key task to allocate each battery string to only one other system component at a time. For the prototype system at hand, no energy management strategies have been developed or explored in the literature.

Hence, the present work breaks fresh ground by proposing a control approach for this novel system design. The contributions of this paper are summarised as follows:
\begin{enumerate}[1)]
    \item A heuristic energy management system with two degrees of control is introduced: First, allocating the battery strings to the other system components, and second, managing the energy exchange with the local distribution grid.
    \item A basic control strategy is compared to an enhanced strategy that includes PV forecasts in the decision making.
    \item An electro-thermal model of the system is used as a framework to test the proposed concepts, using actual PV measurements from the demonstration site.
    \item The performance of the proposed energy management strategies is extensively assessed by Monte Carlo simulations generating a vast number of scenarios which capture the uncertainty of EV charging.
\end{enumerate}

The remainder of the paper is structured as follows: Section~\ref{sec:system} provides an overview of the system under investigation. Section~\ref{sec:method} presents the model, the applied energy management strategies, as well as the case study. Section~\ref{sec:results} summarises and discusses the numerical results, followed by concluding remarks in Section~\ref{sec:conclusion}.
 
 \begin{figure*}[htp]
    \centering
    \includegraphics[]{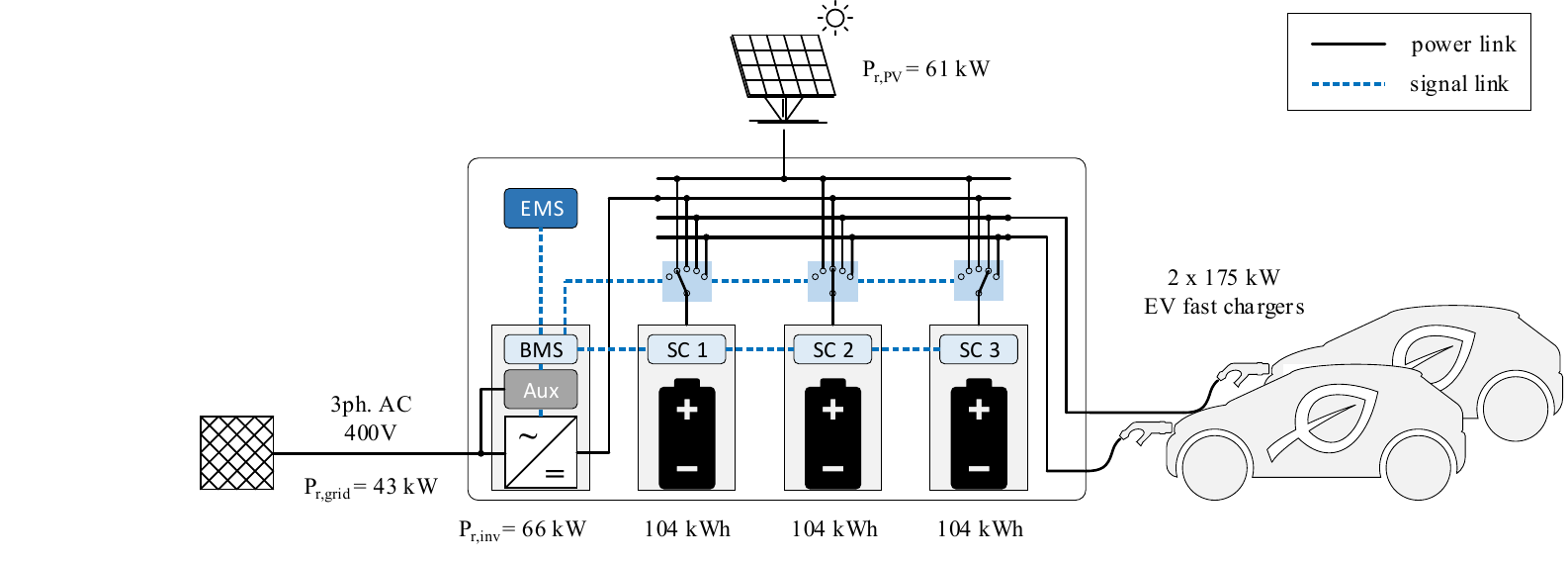}
    \caption{Overview of the hybrid system comprising a multi-battery storage system, a grid-tie inverter, a PV installation, and two EV fast chargers.}
    \label{fig:system_overview}
\end{figure*}

\section{System Overview}\label{sec:system}

Fig.~\ref{fig:system_overview} provides a high-level overview of the hybrid system, comprising the BESS, two 175\,kW EV fast chargers, a 61\,kWp PV system, and a 66\,kW grid-tied inverter. The battery, as a power and energy buffer, enables full usage of the power capability of both the PV system and EV chargers, despite the grid capacity being limited to only 43\,kW. The battery system is hence able to facilitate high power EV charging at two outlets up to 175\,kW at a significantly lower grid rating. The grid connection point is formed by the AC side of the inverter and the auxiliary (Aux). The auxiliary includes the supply of all electronic devices, as well as the thermal management system comprising a heating, ventilation, and air conditioning system. The energy storage consists of $n_{\text{str}} = 3$ battery strings of 104\,kWh each, which can be independently allocated to any of the power components of the system. Each of the strings is equipped with a string controller (SC), monitoring the cell states and managing the cell configuration to meet the voltage requirements of the connected component. Relevant system states, such as the state of energy (SOE) of the string, are transmitted to the battery management system (BMS). The BMS is the central control unit of the BESS responsible for communicating with other units in the container. It further serves as protection system and disconnects strings if they are operated outside their safe operating area. 
 
The task of managing the stored energy in order to fulfil the needs of all system components is taken by the EMS. On the one hand, the EMS must ensure that the charge levels of the strings are high enough to allow also for consecutive charging of EVs. On the other, it must take action if the SOE becomes so high that PV energy could not be captured anymore. To achieve this, the EMS has two degrees of control. First, by allocating the battery strings to respective power components, and second, by managing the energy exchange with the local distribution grid. The EMS communicates its decisions to the BMS, which updates the inverter setpoint and the connection states of the strings.

\section{Methodology}\label{sec:method}

The modelling of the hybrid system comprises two layers. First, the physical model of the power and energy flows in the DC microgrid is presented. Second, the rationale of the energy management system on how to calculate the inverter setpoint and the string allocation is introduced.

\subsection{Physical Multi-Battery Model}

The system shown in Fig.~\ref{fig:system_overview} is modelled on a power and energy level in Matlab Simulink to create a framework for developing and testing EMS strategies. The sign convention of the power flows is defined so that consumption (EV charging, grid export, losses, auxiliary consumption) is represented by positive power values, and energy intake (PV generation, grid import) is represented by negative power values. Hence, the SOE of a string is calculated as 

\begin{equation}
\small
    SOE_{\text{str}} = -\frac{1}{E_{\text{cap,str}}}\int P_{\text{str}}(t)\,dt + SOE_{\text{str,init}},
\end{equation}

where $P_{\text{str}}(t)$ is the string power, $E_{\text{cap,str}}$ the nominal energy capacity of the string, and $SOE_{\text{str,init}}$ the initial charge level at the beginning of the simulation. The strings are operated between SOE limits of 10\% and 90\% to avoid under- and overcharging and stay within their linear voltage region. When crossing these limits, the BMS disconnects the string from the current component, regardless of what the EMS requests. 

Power losses during charging and discharging of reconfigurable battery strings were derived in a previous study \cite{Engelhardt.2021}. In the present model, the efficiency was considered as a linear dependency with 100\% at 0\,kW, and 90\,\% at $\pm$\,120\,kW. The inverter is modelled with an average power efficiency of 98\,\% and a constant standby power of 100\,W, as part of the auxiliary consumption. The power losses of battery and inverter, as well as all auxiliary components, are used as input for a thermal model of the BESS. The thermal model was adopted from a previous work \cite{JesperGejlLage.2021}, and estimates the temperatures of the components inside the container based on their thermal properties, all heat loads, the effects of the cooling system, and the outside air temperature. While the thermal conditions have no particular focus in the present study, they are used to estimate the power consumption of the thermal management system as part of the auxiliary consumption.

The power of the PV system and the two EV chargers are considered as time series inputs, which will be further addressed in Section~\ref{sec:casestudy}. The power at the grid connection point is calculated as 
\begin{equation}\label{eq:grid_contraints}
\small
    P_{\text{grid}}(t) = P_{\text{inv}}(t) - P_{\text{aux}}(t),
\end{equation}
where $P_{\text{inv}}(t)$ is the inverter and $P_{\text{aux}}(t)$ the auxiliary power.

\subsection{Energy Management System}
The EMS is responsible for controlling the energy exchange with the grid and for allocating the three strings to the different power components (EVs, PV, inverter) of the hybrid system. In the following, heuristic control concepts for each of the tasks are proposed.

\subsubsection{Inverter Setpoint}
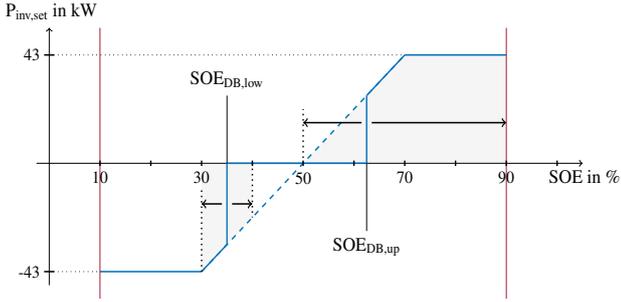
\begin{figure}
\centering
\resizebox{8.7cm}{4cm}{%
\begin{tikzpicture}

    \definecolor{matlab_blue}{rgb}{0,0.4470,0.7410}
    \definecolor{matlab_darkred}{rgb}{0.6350,0.0780,0.1840}

    \coordinate (x30) at (3,2.5);
    \coordinate (x40) at (4,2.5);
    \coordinate (x50) at (5,2.5);
    \coordinate (x85) at (8.5,2.5);
    \coordinate (droop_x30) at (3,0.5);
    \coordinate (droop_x70) at (7,4.5);
    \coordinate (db_low) at (3.5,2.5);
    \coordinate (db_low_pow) at (3.5,1);
    \coordinate (db_high) at (6.25,2.5);
    \coordinate (db_high_pow) at (6.25,3.75);
    
    
    \draw[thin,->] (-0.25,2.5) -- (10.5,2.5) node[rectangle,anchor=north] {
        \renewcommand{\arraystretch}{0.65}
        \begin{tabular}{c} 
            \normalsize SOE in \%
        \end{tabular}};
    
    \draw[thin,->] (0,0.25) -- (0,5) node[rectangle,anchor=south] {
        \renewcommand{\arraystretch}{0.65}
        \begin{tabular}{c} 
            \normalsize P\textsubscript{inv,set} in kW
        \end{tabular}};
    
    \filldraw[draw=black, fill=gray!80,draw opacity=0,fill opacity=0.1] (5,2.5) -- (9,2.5) -- (9,4.5) -- (7,4.5) -- cycle;
    
    \draw[thick,matlab_blue,dashed] (droop_x30) -- (droop_x70);
    \draw[thick,matlab_blue] (1,0.5) -- (droop_x30);
    \draw[thick,matlab_blue] (droop_x70) -- (9,4.5);
    
    \draw[thick,matlab_blue] (db_low) -- (db_high);
    \draw[thick,matlab_blue] (db_high) -- (db_high_pow);
    \draw[thick,matlab_blue] (db_high_pow) -- (droop_x70);
    
    \draw[thick,matlab_blue] (db_low) -- (db_low_pow);
    \draw[thick,matlab_blue] (db_low_pow) -- (droop_x30);
    
    \draw[thick,dotted] (5,2.5) -- (5,3.5);
    \draw[thick,<-] (5,3.25) -- (6.15,3.25);
    \draw[thick,->] (6.35,3.25) -- (9,3.25);
    \draw[very thin] (6.25,2.5) -- (6.25,1.25) node[rectangle,anchor=north] {\normalsize SOE\textsubscript{DB,up}};
    
    \filldraw[draw=black, fill=gray!80,draw opacity=0,fill opacity=0.1] (x30) -- (4,2.5) -- (4,1.5) -- (3,0.5) -- cycle;
    
    \draw[thick,dotted] (3,2) -- (3,0.5);
    \draw[thick,dotted] (x40) -- (4,1.5);
    \draw[thick,<-] (3,1.75) -- (3.4,1.75);
    \draw[thick,->] (3.6,1.75) -- (4,1.75);
    \draw[very thin] (3.5,2.5) -- (3.5,3.75) node[rectangle,anchor=south] {\normalsize SOE\textsubscript{DB,low}};
    
    \draw[ultra thin,matlab_darkred] (1,0) -- (1,5);
    \draw[ultra thin,matlab_darkred] (9,0) -- (9,5);
    \draw[thin,dotted] (0,4.5) -- (droop_x70);
    \draw[thin,dotted] (0,0.5) -- (1,0.5);
    
    \draw[thick] (1,2.55) -- (1,2.45) node[pos=0.75, anchor=north] {\small 10};;
    \draw[thick] (2,2.55) -- (2,2.45);
    \draw[thick] (3,2.55) -- (3,2.45) node[pos=0.75, anchor=north] {\small 30};
    \draw[thick] (4,2.55) -- (4,2.45);
    \draw[thick] (5,2.55) -- (5,2.45) node[pos=0.75, anchor=north] {\small 50};
    \draw[thick] (6,2.55) -- (6,2.45);
    \draw[thick] (7,2.55) -- (7,2.45) node[pos=0.75, anchor=north] {\small 70};
    \draw[thick] (8,2.55) -- (8,2.45);
    \draw[thick] (9,2.55) -- (9,2.45) node[pos=0.75, anchor=north] {\small 90};
    \draw[thick] (10,2.55) -- (10,2.45);
    
    \draw[thick] (-0.1,4.5) -- (0.1,4.5) node[pos=0.25, anchor=east] {\small 43};
    \draw[thick] (-0.1,0.5) -- (0.1,0.5) node[pos=0.25, anchor=east] { \small -43};
    
\end{tikzpicture}
}
\caption{Basic droop control for inverter power based on battery SOE. The enhanced control introduces dynamic deadbands depending on PV forecasts.}
\label{fig:droop}
\end{figure}

The inverter is used to actively control the battery SOE by exporting or importing energy to or from the grid. We compare two different strategies used for the inverter setpoint, which we will in the following call ``base'' and ``enhanced''. In the base strategy, the inverter setpoint is determined by a symmetric droop control based on the overall battery SOE (average string SOE), as shown in Fig.~\ref{fig:droop}. The reference SOE of the droop is 50\,\%, and minimum and maximum grid power is reached at 30\,\% and 70\,\%, respectively. The ``enhanced'' control is based on the same droop characteristic, but introduces flexible upper and lower deadbands that avoid grid exchange around the reference SOE. These deadbands are adapted dynamically according to a one-hour ($t_{\text{PV,fc}} = \SI{1}{h}$) rolling energy forecast of the PV production, $E_{\text{PV,fc}}$. The upper SOE deadband aims at initiating a battery discharge if the forthcoming PV production will bring the battery to its upper charge limit, and is calculated as

\begin{equation}
\small
    SOE_{\text{DB,up}} = 90\,\% - \frac{E_{\text{PV,fc}}}{n_{\text{str}} \cdot E_{\text{cap,str}}} \cdot \frac{P_{\text{r,PV}}}{P_{\text{r,grid}}} \in [50\,\%, 90\,\%],
\end{equation}

where the factor $\frac{P_{\text{r,PV}}}{P_{\text{r,grid}}}$ considers the different power ratings of PV system and grid. The lower SOE deadband aims at avoiding energy import from the grid, in case of forthcoming PV production, and is calculated as

\begin{equation}
\small
    SOE_{\text{DB,low}} = 40\,\% - \frac{40\,\% - 30\,\%}{P_{\text{r,grid}}\cdot t_{\text{PV,fc}}} \cdot E_{\text{PV,fc}} \in [30\,\%, 40\,\%].
\end{equation}

The lower deadband has a minimum value of 30\,\% to retain a sufficient energy content for EV charging, and a maximum value of 40\,\% to avoid unnecessary recharging when being in idle mode (no PV, EVs). The proposed inverter power is communicated from the EMS to the BMS. When updating the power setpoint of the inverter, the BMS considers the current auxiliary consumption to fulfil grid constraints \eqref{eq:grid_contraints}.

\begin{algorithm}[b!]
\caption{Logic of battery string allocation}\label{alg:opt_allocation}
\KwInput{Current string allocation, string SOEs, EV connection states, PV production}
\KwOutput{Proposed string allocation}
~\\

Identification of \emph{available} strings, i.e. strings where no EV is connected \\

\If{EV just unplugged from charger}{
    Make corresponding string available
}

\If{EV just plugged in at charger}{
    Assign string with highest SOE to EV
}

\eIf{no PV production available}{
    
      Assign string to inverter, which is most responsible for battery SOE deviating from 50\,\%: \\
      max((SOE\textsubscript{str} -- 50\,\%) $\cdot$ sgn(SOE\textsubscript{batt} -- 50\,\%))
    }{ 
    \eIf{2 EVs charging}{
        Assign remaining string to PV
    }{
    \eIf{PV production greater than grid capacity}{
        Assign string with lowest SOE to PV\\
        From the remaining strings, assign string to inverter which is most responsible for battery SOE deviating from 50\,\% 
    }{
        Assign string to inverter which is most responsible for battery SOE deviating from 50\,\% \\
        From the remaining strings, assign string with lowest SOE to PV 
    }
    }
    }
\end{algorithm}

\subsubsection{String Allocation}
The technical constraints of the system must be taken into account when designing an algorithm for the allocation of strings and units. Technically, the system only allows to connect EVs, PV, and inverter to battery strings and not directly among each other. This means that e.g. PV power cannot be directly exported to the grid, but has to be stored in a string first. This has to be considered in the decision making, since PV power can only be utilised if the connected string has chargeable capacity. Besides technical constraints, additional considerations are needed: PV production must be always harvested; incoming EVs must always be charged. To ensure a successful EV charging session, the string serving the EV must not be disconnected. Algorithm~\ref{alg:opt_allocation} describes the procedure for allocating the strings, considering the aforementioned constraints. A re-calculation of the string allocation is initiated by the following triggers:

\begin{enumerate}
    \item An EV connects or disconnects at one of the chargers.
    \item The PV system starts/stops producing (e.g. morning/evening or due to maintenance).
    \item Timer of 10 min that checks if a string has an SOE of \textgreater\,80\,\% or \textless\,20\,\%. 
    \item One of the strings reaches its operational limits.
\end{enumerate}

\subsection{Case Study}\label{sec:casestudy}

This section presents the case study for investigating the performance of the two proposed heuristic control strategies under different model configurations. Specifically, the simulation structure, the construction of EV fast-charging profiles, and the origin of PV profiles are introduced.

\paragraph{Simulation Structure}

The simulation model of the DC microgrid has been constructed in Matlab Simulink and is run in a simulation time step of one minute. The model simulates for respective time periods of 14 days in June, September, and November to account for the impact of seasonality of the PV production. Furthermore, different frequencies of EV charging are investigated, ranging from 1 to 30 EVs on average per day.

For assessing the sensitivity in key performance indicators of the two proposed heuristic control strategies, Monte Carlo simulations are performed that capture the uncertainty of EV charging conditional to a certain probability distribution. This is done to extensively test the energy management strategies under random conditions, and not with customised EV charging profiles that might favour the proposed control. For the creation of randomised charging patterns, three dimensions of input variations are considered: for each control strategy, each of the three seasons, and each penetration level of EV charging, 10,000 simulation runs are performed for a two-week time horizon. To this end, the Monte Carlo simulations allow for testing the performance of the proposed control with a vast number of possible scenarios, including also rare and extreme EV charging events. This approach substantiates conclusions based on results with statistical significance.

\paragraph{EV Fast-Charging Profiles}

\begin{figure}[h]
    \centering
    \includegraphics{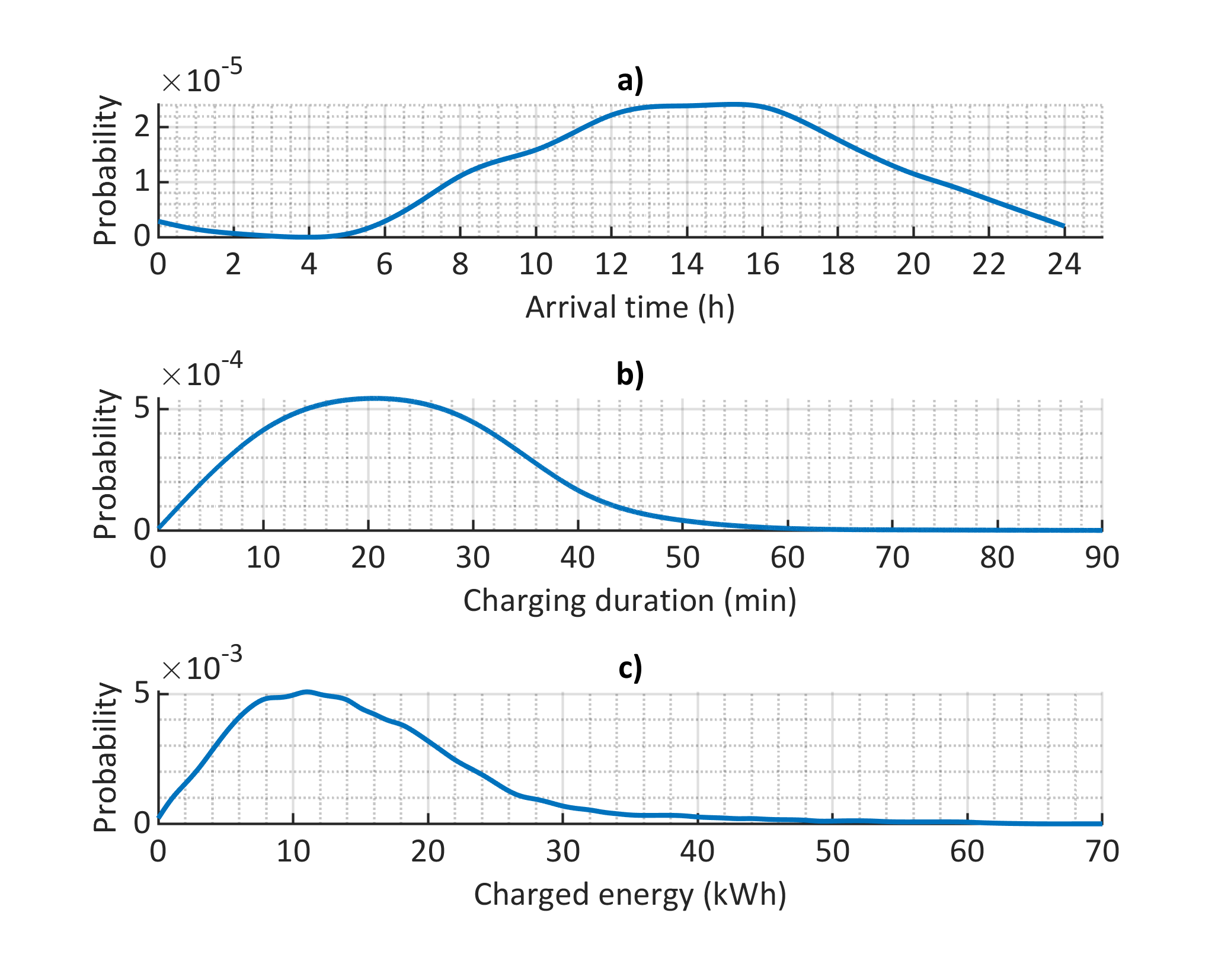}
    \caption{Input probability distribution functions for a) arrival time, b) charging duration, and c) charged energy at fast charging station up to 175\,kW, based on data presented in \cite{Wolbertus.2019,Wolbertus.2020}.}
    \label{fig:input_distributions}
\end{figure}

\begin{figure}[h]
    \centering
    \includegraphics{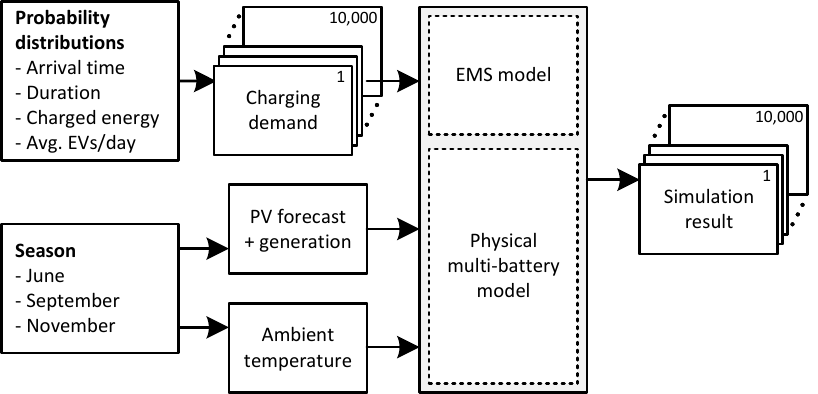}
    \caption{Overview of the simulation structure. For each control strategy, season, and average number of EVs/day, the EMS is tested with 10,000 Monte Carlo runs that capture the uncertainty of EV charging. For each run, the charging demand at the two fast-chargers is randomly generated using probability distributions for the charging behaviour.}
    \label{fig:sim_structure}
\end{figure}

The EV fast-charging profiles are created based on probability distributions of the arrival time, charging duration and charged energy (Figure~\ref{fig:input_distributions}). Data behind these distributions were collected from EV fast-charging stations in the Netherlands with charging powers of up to 175\,kW \cite{Wolbertus.2019,Wolbertus.2020}. The authors processed and analysed data from MultiTankCard, a service provide for mobility in the Netherlands. The published distribution are based on in total over 1 million charging sessions in 2019, providing a strong statistical representation of the charging behaviour at fast-charging stations. Based on the hourly data points provided by the above-mentioned publications, the present study used spline fits to obtain the continuous probability distributions depicted in Figure~\ref{fig:input_distributions}. By drawing from these distributions, it is possible to derive random average charging power profiles with different arrival and connection times. In doing so, no dependency of the charging duration and the charged energy is assumed. For cases where charging powers \textgreater\,175\,kW emerge, new samples are taken from these distributions. Figure~\ref{fig:sim_structure} shows how the EV charging behaviour is integrated in the simulation structure. For each of the 10,000 Monte Carlo runs, the depicted probability distributions are employed to generate random EV power profiles for the two fast-chargers. 

\paragraph{PV Profiles}

The PV system connected to the installed DC microgrid is composed of 10 strings with 23 PV modules. The modules are of the kind Trina TSM-265 with a rated peak power of 265\,W. Since the installation of the hybrid system in June 2021, real PV power measurements for June and September are obtained at two-second time intervals. Subsequently, these profiles were downsampled to a one-minute resolution in order to match the time steps of the simulation. Data from the PV system before June 2021 are only available in hourly averages. Hence, these were used without loss of generality for an analysis of the system in November. 

\begin{figure}[b!]
    \centering
    \includegraphics[width=\columnwidth]{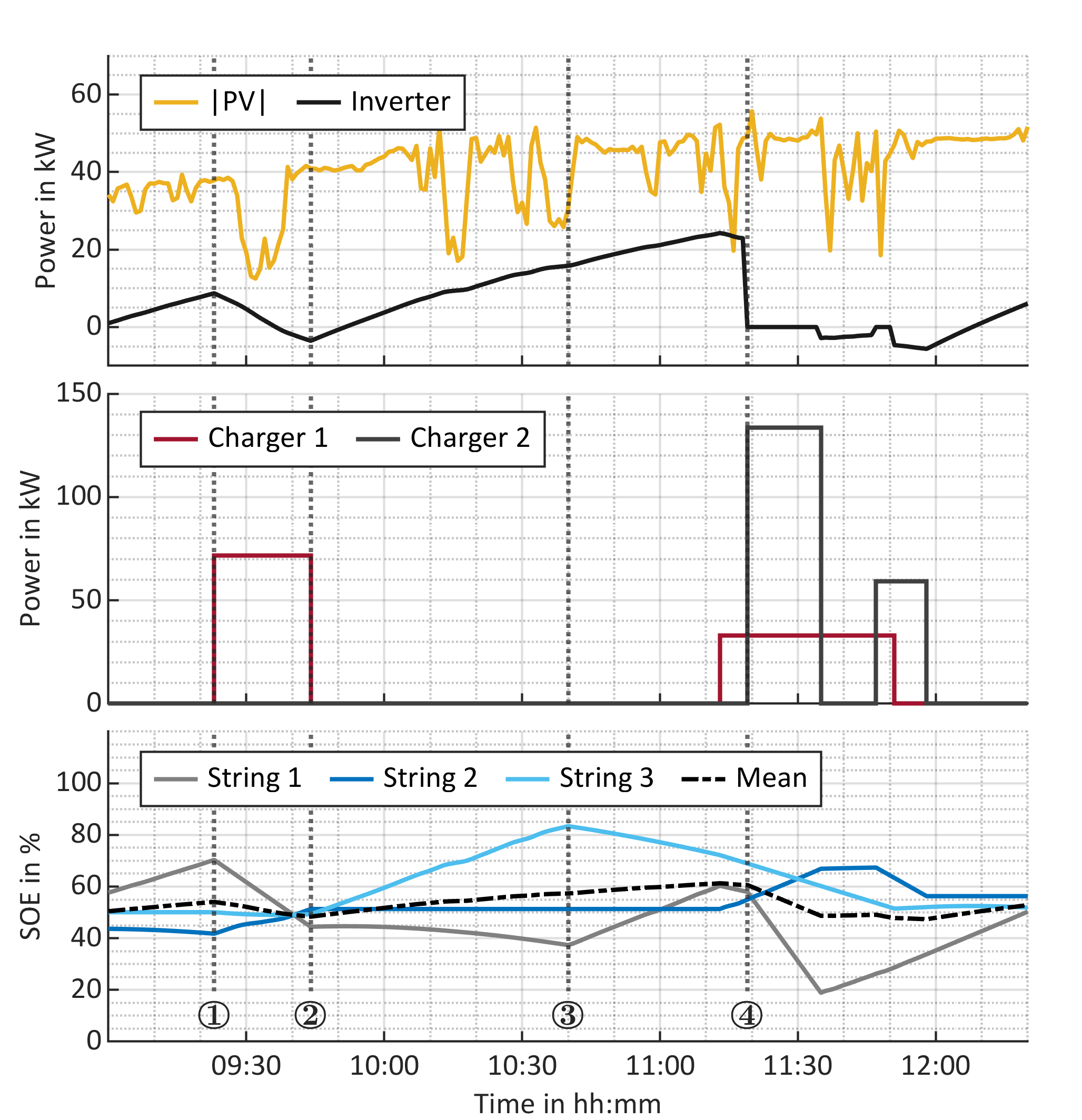}
    \caption{Excerpt of the string allocation for the morning of 8\textsuperscript{th} June 2021.}
    \label{fig:string_allocation}
\end{figure}

\begin{figure*}[t!]
    \centering
    \includegraphics[width=\textwidth]{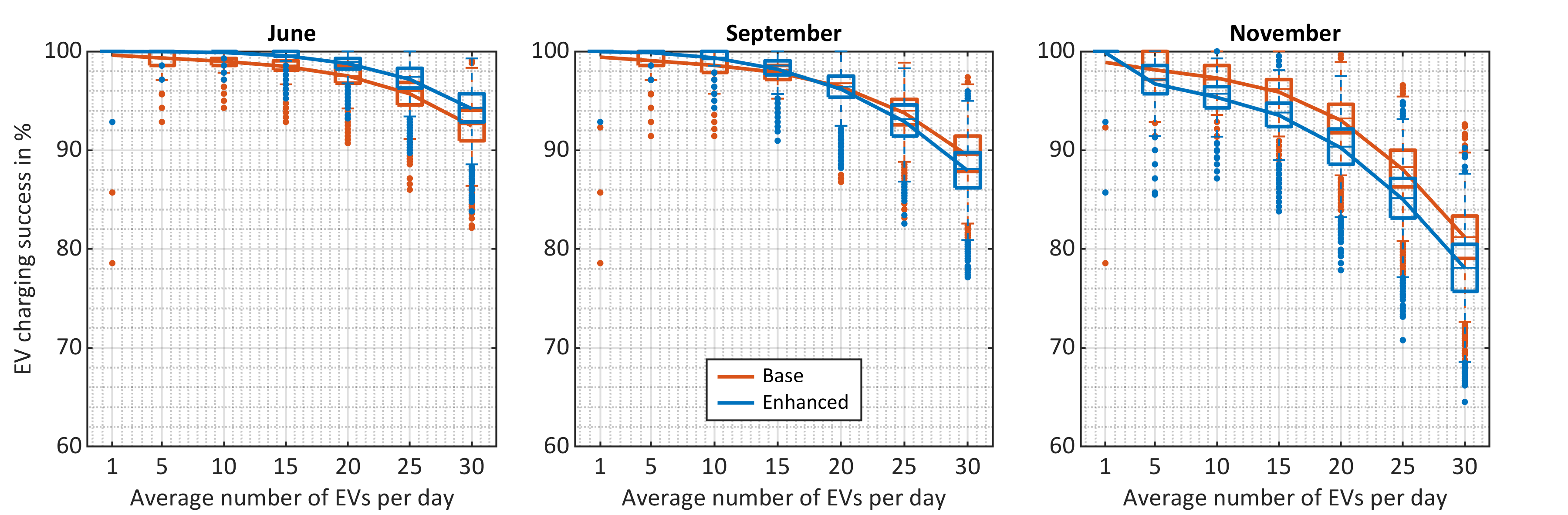}
    \caption{EV charging success rate for the base and enhanced control for varying EV charging frequencies and different seasons.}\label{fig:ev_success}
\end{figure*} 

\begin{figure*}[b!]
    \centering
    \includegraphics[width=\textwidth]{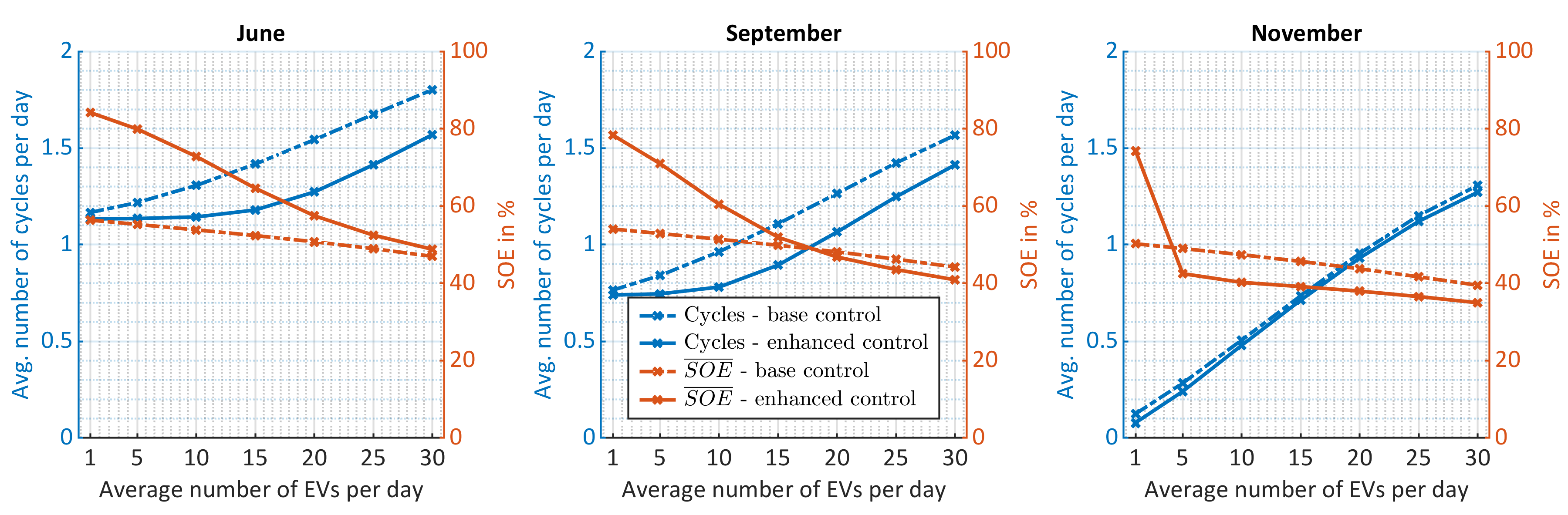}
    \caption{Impact of base and enhanced control on number of cycles and mean battery SOE for varying EV charging frequencies and different seasons.}
    \label{fig:battery_impact}
\end{figure*}

\section{Results and Discussion}\label{sec:results}

This section presents and discusses the numerical results that were obtained from the performed Monte Carlo simulations. In particular, it presents the functioning of the string allocation, the EV charging satisfaction, the battery usage, and key indicators of the system performance.

\subsection{String Allocation}

Fig.~\ref{fig:string_allocation} demonstrates the string allocation during one of the Monte Carlo runs of the base control for an exemplary time window on the 8\textsuperscript{th} June 2021. The plot shows the power profiles of PV and inverter in the top plot, of the two EV chargers in the middle plot, as well as the SOE progression of the battery strings in the lower plot. During the given time window, four EVs are charging with average power values of up to 133.56\,kW. In contrast, the inverter power varies only between -5.63\,kW (import) and 24.25\,kW (export), underlining the buffering effect of the BESS on the grid. 

The figure further exemplifies the functioning of Algorithm~\ref{alg:opt_allocation}, showcasing a set of exemplary triggers for the re-allocation of the battery strings. At time \raisebox{.5pt}{\textcircled{\raisebox{-.9pt} {1}}} an EV connects to Charger~1. The string with highest SOE (String~1) is consequently assigned to that charger, while the string with lowest SOE (String~2) connects to the PV, and the remaining string (String~3) links to the inverter. At time \raisebox{.5pt}{\textcircled{\raisebox{-.9pt} {2}}}, the EV disconnects which causes in turn a re-allocation of the strings, due to changed SOE levels. The string most responsible for mean SOE deviation is now connected to the inverter, while the lowest SOE string is assigned to the PV. At time \raisebox{.5pt}{\textcircled{\raisebox{-.9pt} {3}}}, String~3 is at SOE levels of above 80\,\%. Hence, the control reacts in order to balance the SOEs of the battery strings. String~3, previously connected to the PV, is consequently assigned to the inverter, and String~1 connects to the PV. Time \raisebox{.5pt}{\textcircled{\raisebox{-.9pt} {4}}} serves as an example of an EV connecting at Charger~2, because Charger~1 is already taken by another EV. At that time, String~3 is hence not available. As a consequence, String~1 as the remaining string with the highest SOE switches its connection from the inverter to Charger~2, while String~2 remains connected to the PV system. With two EVs connected during daylight hours, the inverter has no option of being connected due to the premise that PV production must always be used.

\begin{figure*}[t!]
    \centering
    \includegraphics[width=\textwidth]{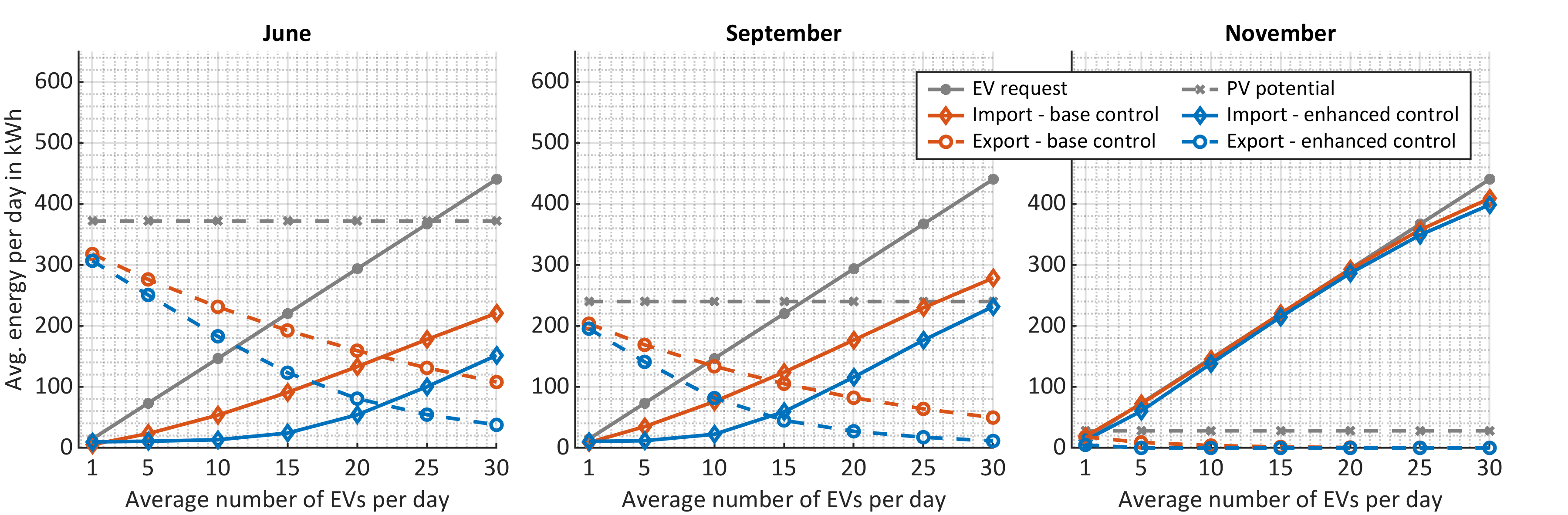}
    \caption{Daily average values for grid import and export for the base and enhanced control, as well as PV energy, and EV charging demand for varying EV charging frequencies and different seasons.}
    \label{fig:energy_grid_exchange}
\end{figure*}

\begin{figure*}[t!]
    \centering
    \includegraphics[width=\textwidth]{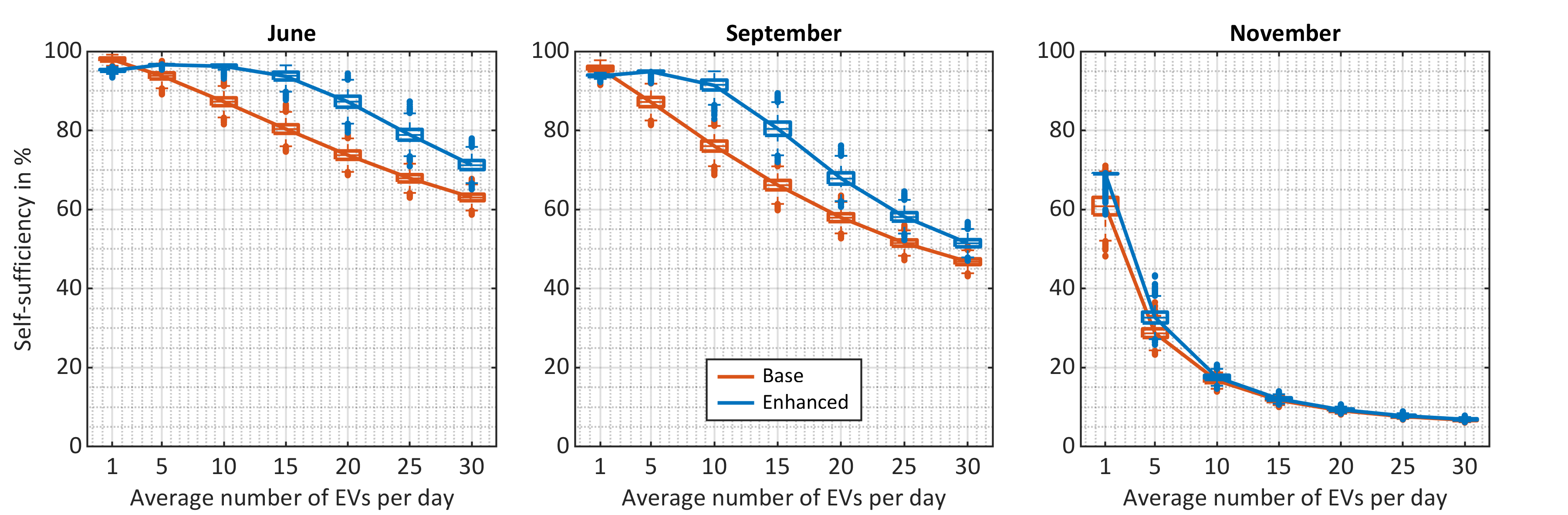}
    \caption{Self-sufficiency for the base and enhanced energy management strategy for varying average numbers of EVs per day and different seasons.}
    \label{fig:self_sufficiency}
\end{figure*}

\subsection{EV Charging}

Fig.~\ref{fig:ev_success} visualises the EV charging success rate as a result of the Monte Carlo simulations comparing the two strategies with a varying number of EV charging instances ranging from 1 to 30 EVs on average per day. An EV charging session is considered as successful if the charger provided \textgreater\,99\,\% of the requested energy. Although this is a strict criterion for defining charging success, it gives an indication of the lower bound of the overall success rate for EV charging. For each Monte Carlo run of two weeks, a success rate of EV charging is calculated. The success rates for each strategy and EVs/day ratio are summarised in individual box plots. Each box plot is based on 10,000 two-week simulation runs. The boxes indicate the interquartile ranges, while the whiskers are set to 1.5 times the respective quartiles. The lines connect the mean values of the distributions. In June, the enhanced control enables EV charging with a high success rate that decreases to 94.20\,\% if 30 EVs use the chargers each day. In comparison, the base control reaches an average success rate of 92.45\,\% for the same number of EVs. In September, the enhanced control performs better to the threshold of 15 EVs/day, while the base control achieves a higher success rate beyond that point. In November, the base control outperforms the enhanced method, although both strategies have lower success rates than during the other months. The progression of the success rates can be explained by the average battery SOE for the different scenarios, as compared in Fig.~\ref{fig:battery_impact}. With a higher charge level, consecutive EV charging sessions are more likely to be provided with the requested energy. Hence, when comparing the average battery SOEs of the two strategies, they show the same relation as the EV success rate. Moreover, the premise of keeping strings and EV connected will always lead to a failed (not completed) charging session if the requested energy is larger than the energy content of the respective string -- regardless of the overall SOE of the BESS. Hence, methods should be reviewed that allow pausing the power transfer without terminating the entire charging process, in order to swap the discharged string. This is expected to improve the success rate significantly. It should be noted that with the applied definition of success rate of \textgreater\,99\,\% of the requested energy delivered, a 95\,\% service delivery still counts as a failure. Customers of fast-charging stations might define charging success less strict and hence be content with slightly lower delivered energy than requested.

\subsection{Battery Usage}

Capacity degradation of lithium-ion batteries is commonly determined by two effects: cycle and calendar ageing. While cycle ageing occurs when a battery is charged and discharged, calendar ageing also takes place when a battery is unused, and is affected by battery SOE and temperature \cite{Barre.2013}. Hence, the number of full cycles and the average battery SOE are important indicators for the battery usage and as such a suitable metric for evaluating EMS performance. As seen in Fig.~\ref{fig:battery_impact}, the base control entails higher numbers of cycles during all months, compared to the enhanced control. In contrast, the enhanced control allows a higher SOE in summer and a lower SOE in winter due to the flexible deadband. Consequently, the two control strategies show opposing relations for cycle and calendar ageing. It should be noted that some studies show that average charge levels of \textgreater\,70\,\% accelerate calendar ageing significantly \cite{Keil.2016}. Hence, the enhanced control could be adjusted to decrease the upper deadband in periods with little EV charging. However, long-term studies are necessary to confirm the impact of the charge levels on the battery degradation.

\subsection{System Performance}

To assess the buffering effect of the battery storage, the energy flows to and from the hybrid system can be compared. Fig.~\ref{fig:energy_grid_exchange} illustrates daily average values for grid import and export, PV energy, and EV charging demand, as a result of the performed Monte Carlo simulations. The dashed grey line represents the potential PV energy in the respective seasons, while the solid grey line depicts the increase in EV charging demand with increasing number of EVs/day. For June and September, the enhanced control entails significantly less grid exchange, i.e. import and export, due to the incorporated deadband that requires less immediate control action. For 30 EVs/day in June, the import and export from and to the grid can be decreased compared to the base control by 65\,\% and 31.3\,\%, respectively. In November, the PV production level is insignificant. Hence, the import in both the base and enhanced control strategies align increasingly with the requested EV energy. The lower EV charging success rate for high numbers of EVs/day in November is also reflected in decreasing grid import in both control strategies. The premise to always utilise PV energy when available prevents the use of the inverter to import energy from the grid, in case two EVs charge simultaneously. Hence, in months with low PV potential this might cause a depletion of the battery and, consequently, a lower success rate of EV charging.

Based on grid exchange, PV production, and EV consumption, the self-sufficiency of the hybrid system can be estimated. It is defined as the percentage of EV consumption that was covered by the local PV production, taking into account the power losses of strings and inverter, as well as the auxiliary consumption. Fig.~\ref{fig:self_sufficiency} shows the distributions of obtained self-sufficiency values for different average numbers of EVs per day. For both control strategies, the self-sufficiency generally decreases with an increasing number of EVs per day. The enhanced control strategy, incorporating a dynamic deadband based on the rolling PV forecast, obtains higher self-sufficiency values for the month of June and September. In June, even with an average of 30 EV charging instances per day, the self-sufficiency ends for most of the simulation runs above 70\,\% in case of enhanced control. In November, due to low PV production, the two control strategies equalise in terms of self-sufficiency as both are equivalently reliant on grid import to fulfil EV charging. 

\subsection{Future Considerations}

The presented EMS can be further improved by considering additional inputs in the decision making. Including energy prices for importing and exporting energy introduces economical considerations with the aim of reducing EV charging costs. Utilising historical EV charging behaviour on the demonstration site allows for an improved management of the battery SOE to improve the EV charging success rate. Furthermore, the battery SOE can be included in the decision making in order to reduce calendar ageing. The strict constraint of always utilising local PV production when available could be relaxed for achieving higher charging success rates by prioritising grid imports when PV production is low. Formulating these tasks as a multi-objective optimisation problem is one possible way to capture the above mentioned points. In this regard, the presented heuristic control can serve as a benchmark to assess the performance of the optimisation.

\section{Conclusion}\label{sec:conclusion}

Hybrid fast-charging stations with incorporated battery storage and local renewable generation units can foster the decarbonisation of the transport sector, while reducing the grid impact on distribution level. The present study focuses its attention on a novel system architecture comprising a multi-string battery system with reconfigurable topology that allows the direct coupling to other DC components through a busbar matrix. This paper breaks fresh ground by proposing an energy management system for this prototype battery storage interfacing a PV installation, a low-rated distribution grid connection, and two high power EV fast-chargers. The key task of the energy management system is to allocate each of the three battery strings to only one of the other four components at a time. For the energy exchange with the grid, a base control is compared to an enhanced strategy that includes PV forecasts in the decision making. The performance of these proposed strategies is assessed by Monte Carlo simulations with a vast number of EV charging scenarios that allow a statistically robust interpretation of the results. For an average number of 20 EVs per day in summer, the base and enhanced strategy fulfil the energy requests of the EVs in 97.5\,\% and 98.8\,\% of the charging instances, respectively. Yet, the enhanced strategy achieves significantly higher self-sufficiency of the hybrid system at 87.3\,\%, while the base strategy yields 73.8\,\%. In winter with significantly lower PV production, however, the self-sufficiency drops to 9.3\,\% (enhanced) and 9.1\,\% (base). In November,
the base control achieves a higher success rate than the enhanced method, although both strategies still complete 90.2\,\% (enhanced) and 93\,\% (base) of the charging processes, considering that they prioritise low local PV production over grid import. All in all, the enhanced control reduces the number of battery cycles, at the cost of higher average charge levels, compared to the base control. The ongoing demonstration activities within the InsulaE project invite for model improvements based on real EV charging data.

\section*{Acknowledgements}
This work has received funding from the H2020 Insulae project under the Grant Agreement No. 824433, and from the IFD funded TOPChargE project under the Grant Agreement No. 9090-00035A.

\bibliography{bibliography}

\end{document}